\documentclass[aps,prc,superscriptaddress,nofootinbib,notitlepage,showpacs,showkeys]{revtex4-1}
\usepackage{graphicx}  
\usepackage{dcolumn}   
\usepackage{bm}        
\usepackage{amssymb}   
\usepackage{amsmath}   
\usepackage{relsize}
\usepackage{xfrac}
\usepackage{color}
\definecolor{dgreen}{cmyk}{1.,0.,1.,0.2}        
\definecolor{orange}{cmyk}{0.,0.353,1.,0.}    

\usepackage[bookmarks]{hyperref}

\def\snn{\mbox{$\sqrt{s_{_{\rm NN}}}$}}

\newcommand{\bra}[1]{\langle #1|}
\newcommand{\ket}[1]{|#1\rangle}
\newcommand{\braket}[2]{\langle #1|#2\rangle}

\newcommand{\di}{{\rm d}}
\newcommand{\ii}{i}

\def\wT{{\widehat T}}

\def\wJ{{\widehat J}}

\def\wP{{\widehat P}}

\def\wLa{{\widehat \Lambda}}

\def\wrho{{\widehat{\rho}}}

\newcommand{\tr}{{\rm tr}}  
  
\newcommand{\e}{{\rm e}}

\newcommand{\p}{{\rm p}}

\newcommand{\be}{\begin{equation}}
\newcommand{\ee}{\end{equation}}                                                                               
\newcommand{\bea}{\begin{eqnarray}}
\newcommand{\eea}{\end{eqnarray}} 
\begin{document}
	

\title{Polarization transfer in hyperon decays and its effect in relativistic nuclear collisions}
\author{Francesco Becattini}
\altaffiliation{becattini@fi.infn.it}
\affiliation{Dipartimento di Fisica,
	Universit\`{a} di Firenze, and INFN, Sezione di Firenze, Florence,
	Italy} 
\author{Gaoqing Cao}
\altaffiliation{caogaoqing@mail.sysu.edu.cn} 
\affiliation{ School of Physics and Astronomy, Sun Yat-Sen 
	University, Guangzhou 510275, China.} 
\author{Enrico Speranza}
\altaffiliation{esperanza@itp.uni-frankfurt.de}
\affiliation{Institute for Theoretical Physics, Goethe University, 
	D-60438 Frankfurt am Main, Germany.} 

\begin{abstract}
We calculate the contribution to the polarization of $\Lambda$ hyperons in relativistic
nuclear collisions at high energy from the decays of $\Sigma^*(1385)$ and $\Sigma^0$, which
are the predominant sources of $\Lambda$ production besides the primary component, as a
function of the $\Lambda$ momentum. Particularly, we estimate the longitudinal component
of the mean spin vector as a function of the azimuthal angle in the transverse plane,
assuming that primary $\Sigma^*$ and $\Sigma^0$ polarization follow the predictions of local 
thermodynamic equilibrium in a relativistic fluid. Provided that the rapidity dependence
around midrapidity of polarization is negligible, we find that this component of the
overall spin vector has a very similar pattern to the primary one. Therefore, we conclude 
that the secondary decays cannot account for the discrepancy in sign between experimental 
data and hydrodynamic model predictions of the longitudinal polarization of $\Lambda$
hyperons recently measured by the STAR experiment at RHIC.
\end{abstract}

\pacs{25.75.Ld, 25.75.Gz, 05.70.Fh}

\maketitle

\section{Introduction}
	
The evidence of global polarization of $\Lambda$ hyperons in relativistic nuclear collisions~\cite{starnat}
is having a remarkable impact in this field. Indeed, the global polarization turns out to be 
in a very good quantitative agreement with the combined predictions of thermodynamics and hydrodynamics
~\cite{becacsernai,becavort,karpbeca,xiecsernai}. These predictions are based on the 
assumption that local thermodynamic equilibrium is achieved at some early stage of the process 
(Quark Gluon Plasma - QGP - formation) and maintained until hadronization, where the fluid basically breaks
up into a kinetic hadronic system. Thermodynamics provides that particles at hadronization are polarized
if the thermal vorticity tensor $\varpi$ (see below for definition) is non-vanishing while the hydrodynamic
model predicts $\varpi$ at the hadronization once initial conditions of QGP are set.

The model is very successful for the {\em global} polarization, that is for the mean spin vector of 
the $\Lambda$ hyperon, which is parallel to the overall angular momentum of the colliding nuclei, at 
several energies. This model also predicts the mean spin vector as a function of momentum of the $\Lambda$ hyperon~\cite{becacsernai,becavort,karpbeca,pang2016}. 
Particularly, it was observed that the longitudinal - that is, along the beam line - component of the $\Lambda$
mean spin vector shows a very similar pattern to that of the azimuthal particle spectra, the
so-called elliptic flow~\cite{becavort,xiecsernai0}. This feature and more characteristics
of the longitudinal component of the polarization were analyzed and discussed in refs.~\cite{becakarp,voloshin}.
	
The oscillations of the longitudinal polarization of $\Lambda$ as a function of the azimuthal 
angle have indeed been observed by the STAR experiment~\cite{niida}, yet with a flipped sign
with respect to the thermodynamic-hydrodynamic calculations \cite{becakarp}. Interestingly, the sign 
prediction was confirmed in an AMPT-based calculation of the thermal vorticity pattern at hadronization~\cite{xia}
as well as in a single-freeze out scenario \cite{flork2}. 
Furthermore, the STAR experiment has measured the azimuthal dependence of the component of the 
mean spin vector at \snn = 200 GeV~\cite{star2} along the total angular momentum axis and found 
that it is markedly different from the predictions of the model \cite{becacsernai,becavort,karpbeca,xghuang}, 
exhibiting a maximum along the reaction plane and a minimum in the orthogonal direction.

There might be many reasons for these discrepancies and different options have been lately discussed
in literature: non-equilibrium of spin degrees of freedom and the consequent need of developing a 
spin kinetic theory \cite{kin1,kin2,kin3,kin4,kin5,kin6}; introduction 
of a spin potential breaking the equivalence of stress-energy and spin tensors \cite{flork1,Florkowski:2017dyn2,bfs}; 
final-state hadronic potentials \cite{csernaikapusta} and others. However, before invoking 
alternative theoretical scenarios, it is desirable to consider and possibly rule out the simplest 
mechanisms and the simplest among the simplest - at least conceptually - is resonance decay. It 
should be stressed that the calculations of the polarization pattern in $\Lambda$ momentum space 
in the thermodynamic-hydrodynamic framework were made only for primary particles, namely those 
directly emitted from the hadronizing source. However, most $\Lambda$'s are secondary, i.e., decay 
products of higher lying states and so one wonders whether the contribution of the secondaries 
could modify the polarization pattern of the primaries and account for the experimental observations. 

In a previous work \cite{becalisa}, the decay contribution to global polarization was calculated
for several channels and found to be just a slight cumulative correction to the primary component
\cite{karpbeca}. In this work, we extend this calculation differentially in momentum space, thereby 
answering the above question.
	
The paper is organized as follows: In Sec.~\ref{spindenm}, a formalism is developed for single particle 
spin density matrix and the relevant mean spin vector by assuming no interaction between the final 
hadrons. In Sec.~\ref{ptransfer}, we derive the formalism for reduced spin density matrix and polarization 
transfer in two-body decays, which are then applied to the concrete cases contributing to $\Lambda$ 
polarization: $\Sigma^* \rightarrow \Lambda \pi$ in Sec.~\ref{Sigmas} and $\Sigma^0 \rightarrow \Lambda \gamma$ 
Sec.~\ref{Sigman}. For polarization transfer, momentum average over the Mother particle distribution 
should also be involved, which is presented in Sec.~\ref{sec:momentum}. The numerical results are presented 
in Sec.~\ref{results}, which is followed by brief conclusions and discussions in Sec.~\ref{conclusions}. 

\subsection*{Notations and conventions}
	
In this paper, we use the natural units with $\hbar=c=k_B=1$.

The Minkowskian metric tensor is ${\rm diag}(1,-1,-1,-1)$; for the Levi-Civita
symbol we use the convention $\epsilon^{0123}=1$. Operators in 
Hilbert space will be denoted by a large upper hat, e.g. $\wT$; while unit 
vectors with a small upper hat, e.g. $\hat v$. Vector and tensor products are denoted by ``$\cdot$" 
and ``$:$", respectively, e.g. $b \cdot \wP=b_\mu\wP^\mu$ and $\varpi : \wJ=\varpi_{\mu\nu}\wJ^{\mu\nu}$, 
where summation over repeated indices is understood.

The summation convention is also implemented for the angular momentum component indices: if 
the indices show more than once as superscripts or subscripts in the formula, they should be 
summed over all possible values. For example, we should sum over $m$ in the numerator 
of (\ref{rho1}) and over $m,\lambda_1$ and $\lambda_2$ in the denominator as 
$|D^j(\varphi_*,\theta_*,0)^{m}_{\lambda}|^2=D^j(\varphi_*,\theta_*,0)^{m \,*}_{\lambda_1-\lambda_2}D^j(\varphi_*,\theta_*,0)^{m}_{\lambda_1-\lambda_2}$.

 The following conventions are used for the energy-momentum variables:
 the variables in the Mother particle's rest frame will be labelled by a subscript ``$*$". 
 In general, the four-momentum of the Mother in the laboratory frame is denoted by a capital ``$P$" 
 and with a ``$p$" for the Daughter particles, which will be emphasised further when needed. 
 Finally, the three-momentum is denoted by the roman font, i.e., ``{\rm P}" and ``{\rm p}" for 
 the Mother and Daughter, respectively.

\section{Spin density matrix}\label{spindenm}

A crucial ingredient for our calculation is the spin density matrix at local thermodynamic 
equilibrium for hadrons at particlization stage. The definition of the spin density matrix 
for free particles with four-momentum $p$ in Quantum Field Theory (QFT) is as follows:
\be\label{spindens}
\Theta(p)_{\sigma \sigma'} = \frac{\tr (\wrho \, a^\dagger(p)_{\sigma'} a(p)_{\sigma})}
{\sum_\tau \tr (\wrho \, a^\dagger(p)_{\tau} a(p)_\tau)},
\ee
where $a(p)_\sigma$ are destruction operators of the particle with momentum $p$ in the 
spin state $\sigma$ and $\wrho$ is the density operator representing the state of
the field. As it is known, the meaning of $\sigma$ depends on the choice of the 
so-called {\em standard Lorentz transformation} $[p]$ taking the unit time vector 
into the direction of the four-momentum $p$ of the massive particle~\cite{Moussa}. 
For most common choices, $\sigma$ is either the third component of the spin in the 
rest frame~\cite{Weinberg} or the helicity~\cite{Tung}; (see also ref.~\cite{Chung}) 
we will use the latter throughout, corresponding to the transformation:
\be\label{standard}
[p] = {\sf R}(\varphi,\theta,0) {\sf L}_z(\xi) ={\sf R}_z(\varphi) {\sf R}_y(\theta) {\sf L}_z(\xi), 
\ee
where ${\sf L}_z(\xi)$ is a Lorentz boost along the $z$ axis with hyperbolic angle $\xi$ 
($\cosh \xi = \varepsilon/m$) and ${\sf R}(\varphi,\theta,0)$ is a rotation with the Euler angles
$\varphi,\theta,0$ associated to the spherical coordinates of the momentum ${\bf p}$.

From the spin density matrix (\ref{spindens}), one can readily determine the 
mean spin vector by utlizing the decomposition of this vector operator on the space-like orthonormal 
vector basis $n_i(p) = [p](\hat e_i)$~\cite{Moussa,becalisa} associated with the particle momentum $p$:
\be\label{meanspin}
S^\mu(p) = \sum_{i=1}^3 D^S({\sf J}^i)_{\sigma\sigma'} 
\Theta(p)_{\sigma'\sigma} n_i(p)^\mu = \sum_{i=1}^3 \tr (D^S({\sf J}^i) \Theta(p)) [p](\hat e_i)^\mu
= \sum_{i=1}^3 [p]^\mu_i \tr (D^S({\sf J}^i) \Theta(p)), 
\ee
where ${\sf J}^i$ are the angular momentum generators, $D^S({\sf J})$ their irreducible
representation matrices of spin $S$ and $\hat e_i$ is the $i$-th vector of the orthonormal basis. 
It should be stressed that, in spite of its appearance, the mean spin vector \eqref{meanspin} is 
independent of the particular choice of the standard Lorentz transformation $[p]$. Indeed, the 
spin density matrix \eqref{spindens} also depends on the conventional choice of the standard 
Lorentz transformation -- the definition of the spin variables $\sigma$, and this compensates 
the dependence on the matrix elements $[p]^\mu_i$ (see Appendix~\ref{app:lorentz}).

The angular momentum generators ${\sf J}$ and the boost generators ${\sf K}$ are the vector 
components of the anti-symmetric tensor generators of the Lorentz transformations. In covariant form,
\be\label{dspin}
D^S({\sf J}^\lambda) = -\frac{1}{2} \epsilon^{\lambda\mu\nu\rho} D^S(J_{\mu\nu})\hat t_\rho,
\qquad \qquad  D^S({\sf K}^\lambda) = D^S(J^{\nu\lambda})\hat t_\nu ,
\ee
implying the decomposition:
\be\label{dspin2}
D^S(J_{\mu\nu}) = \epsilon_{\mu\nu\rho\sigma} D^S({\sf J}^\rho)\hat t^\sigma
+ D^S({\sf K}_\nu)\hat t_\mu - D^S({\sf K}_\mu)\hat t_\nu, 
\ee
where $\hat t$ is the unit time vector with components $(1,0,0,0)$, which implies $\lambda \ne 0$
in \eqref{dspin}. Thereby, the Eq.~\eqref{meanspin} can be rewritten with the full range 
of indices:
\be\label{meanspincov}
S^\mu(p) = [p]^\mu_\nu \tr (D^S({\sf J}^\nu) \Theta(p)). 
\ee

The calculation of the spin density matrix \eqref{spindens} for a general spin is not an easy 
task in QFT. Even for the simplest non-trivial case of a density operator involving the angular 
momentum, an exact solution is not known. However, it is possible to find an explicit exact solution 
for single relativistic quantum particles neglecting quantum statistics, i.e. quantum field effects. 
In this case, the general equilibrium density operator $\wrho$ reads:
$$
\wrho = \frac{1}{Z} \exp \left[ -b \cdot \wP + \frac{1}{2} \varpi : \wJ \right],
$$
where $b$ is a constant time-like four-vector and $\varpi$ a constant anti-symmetric tensor which
turns out to be the thermal vorticity \cite{becaprl,becaspin}; $\wP$ and $\wJ$ are the conserved total 
four-momentum and total angular momentum-boosts operators, respectively. Viewing the system 
as a set of non-interacting distinguishable particles, we can write:
$$
\wP = \sum_{i} \wP_i, \qquad \qquad  \wJ = \sum_i \wJ_i,
$$
and consequently,
$$
\wrho = \otimes_i \wrho_i  \qquad \qquad {\rm with} \qquad \wrho_i = 
\frac{1}{Z_i} \exp \left[ -b \cdot \wP_i + \frac{1}{2} \varpi : \wJ_i \right].
$$
By using Poincar\'e group algebra, it can be shown that each $\wrho_i$ can be factorized
as \cite{becaprep}:
$$
\wrho_i = \frac{1}{Z_i} \exp \left[ -\tilde b \cdot \wP_i \right] 
\exp \left[\frac{1}{2} \varpi : \wJ_i\right],
$$
where 
$$
\tilde b_\mu = \sum_{k=0}^\infty \frac{\ii^k}{(k+1)!} \underbrace{\left( 
	\varpi_{\mu\nu_1} \varpi^{\nu_1\nu_2} \ldots \varpi_{\nu_{k-1}\nu_k} \right)}_\text{k times} 
b^{\nu_k}. 
$$
Then, for a single particle with momentum $p$, the spin density matrix can be expressed
as:
\be\label{spindens2}
\Theta(p)_{\sigma \sigma'} = \frac{\bra{p,\sigma} \wrho_i \ket{p,\sigma'}}
{\sum_\tau \bra{p,\tau} \wrho_i \ket{p,\tau}}.
\ee

To derive the explicit form for (\ref{spindens2}), we use an analytic continuation technique: 
we first determine $\Theta(p)$ for imaginary $\varpi$ and then continue the function to real
values. If $\varpi$ is imaginary, $\exp[\varpi:\wJ/2] \equiv \wLa$ is just a unitary 
representation of a Lorentz transformation, and then one can use the well known relations 
in group theory to obtain:
\be\label{spindens3}
\Theta(p)_{\sigma \sigma'} = \frac{\bra{p,\sigma} \wLa \ket{p,\sigma'}}
{\sum_\tau \bra{p,\tau} \wLa \ket{p,\tau}} = 
\frac{2\varepsilon\delta^3({\bf p} - {\bf \Lambda (p)}) W(p)_{\sigma\sigma'}}
{2\varepsilon\delta^3({\bf p} - {\bf \Lambda(p)}) \sum_\tau W(p)_{\tau\tau}}.
\ee
where by ${\bf \Lambda(p)}$ stands for the spacial part of the four-vector ${\sf \Lambda}(p)$.
In (\ref{spindens3}), $W(p)$ is the Wigner rotation, that is:
$$
W(p) = D^S([{\sf \Lambda}p]^{-1} {\sf \Lambda} [p]),
$$
where $D^S$ stands for the finite-dimensional representation of dimension $2S+1$, the
so-called $(0,2S+1)$ representation~\cite{Tung} of the SO(1,3)-SL(2,C) matrices in the argument 
\footnote{Note that the Lorentz transformations in Minkowski space-time and their counterparts 
of the fundamental $(0,1/2)$ representation of the SL(2,C) group are henceforth identified. 
Particularly, the standard Lorentz transformation $[p]$ stands for either a SO(1,3) transformation 
or a SL(2,C) transformation.}.
We have also used the covariant normalization of states:
$$
  \braket{p,\sigma}{p, \sigma'} = 2 \varepsilon \delta^3({\bf p}-{\bf p}')
   \delta_{\sigma\sigma'}.
$$
Altogether, we have:
$$
  \Theta(p)_{\sigma \sigma'} = \frac{D^S([p]^{-1} {\sf \Lambda} [p])_{\sigma\sigma'}}
  {\tr (D^S({\sf \Lambda}))},
$$
which seems to be an appropriate form to be analytically continued to real $\varpi$. 
However, it is not satisfactory yet as the continuation to real $\varpi$, that is
\footnote{We will also use the notation:
$$
\Sigma_S =  D^S(J)  \qquad \qquad {\rm with} \qquad \Sigma_{1/2} \equiv \Sigma.
$$
}
$$
  D^S({\sf \Lambda}) =  \exp\left[- \frac{\ii}{2} \varpi : \Sigma_S\right] 
  \to \exp\left[\frac{1}{2} \varpi : \Sigma_S\right]
$$
does not give rise to a hermitian matrix for $\Theta(p)$ as it should. This problem can 
be fixed by taking into account that $W(p)$ is the representation of a rotation, hence 
unitary. We can thus replace $W(p)$ with $(W(p) + W(p)^{-1\dagger})/2$ in (\ref{spindens3}) 
and, by using the property of of SL(2,C) representations $D^S(A^\dagger) = D^S(A)^\dagger$ 
\cite{Moussa} we obtain:
$$
  \Theta(p) = \frac{D^S([p]^{-1} {\sf \Lambda} [p])+
   D^S([p]^{\dagger} {\sf \Lambda}^{-1 \dagger}[p]^{-1\dagger})}
   {\tr (D^S({\sf \Lambda})+D^S({\sf \Lambda})^{-1 \dagger})},
$$
which will give a hermitian result because the analytic continuation of $\Lambda^{-1\dagger}$ reads:
$$
  D^S({\sf \Lambda}^{-1\dagger}) \to \exp\left[\frac{1}{2} \varpi : \Sigma_S^\dagger\right].
$$
Therefore, the final expression of the spin density matrix is:
\be\label{spindensf}
  \Theta(p) = \frac{D^S([p]^{-1} \exp[(1/2) \varpi : \Sigma_S] [p])+ D^S([p]^{\dagger} 
   \exp[(1/2) \varpi : \Sigma^\dagger_S] [p]^{-1\dagger})}
   {\tr (\exp[(1/2) \varpi : \Sigma_S] + \exp[(1/2) \varpi : \Sigma_S^\dagger])},
\ee
which is manifestly hermitian.

The expression (\ref{spindensf}) can be further simplified. By taking the above matrices as
SO(1,3) transformations and using known relations in group theory, we have
$$
  [p]^{-1} \exp \left[ \frac{1}{2} \varpi : J \right] [p] = \exp\left[ 
   \frac{1}{2} \varpi^{\mu\nu} [p]^{-1} J_{\mu\nu} [p] \right] =  
   \exp\left[ \frac{1}{2} \varpi^{\mu\nu} [p]^{-1\alpha}_\mu [p]^{-1^\beta}_\nu J_{\alpha\beta}
   \right].
$$
We can now apply the Lorentz transformation $[p]$ to the tensor $\varpi$:
\be\label{varpiboost}
  \varpi^{\mu\nu} [p]^{-1\alpha}_\mu [p]^{-1 \beta}_\nu = \varpi^{\alpha\beta}_{*}(p)
\ee
to realize that $\varpi_*^{\alpha\beta}$ are the components of the thermal vorticity tensor 
in the rest-frame of the particle with four-momentum $p$. Note that these components are obtained 
by back-boosting with $[p]$, which is not a pure Lorentz boost in the helicity scheme. Finally, 
(\ref{spindensf}) becomes
\be\label{spindensf2}
  \Theta(p) = \frac{D^S(\exp[(1/2) \varpi_*(p) : \Sigma_S])+ D^S(\exp[(1/2) \varpi_*(p) : \Sigma^\dagger_S])}
   {\tr (\exp[(1/2) \varpi : \Sigma_S] + \exp[(1/2) \varpi : \Sigma_S^\dagger])}.
\ee
The thermal vorticity $\varpi$ is usually $\ll 1$; in this case, the spin density matrix can be
expanded in power series around $\varpi=0$. Taking into account that $\tr(\Sigma_S)=0$, we have:
$$
  \Theta(p)^\sigma_{\sigma'} \simeq \frac{\delta^\sigma_{\sigma'}}{2S+1} + \frac{1}{4(2S+1)} 
  \varpi_*(p)^{\alpha\beta}(\Sigma_{S{\alpha\beta}}+\Sigma_{S{\alpha\beta}}^\dagger)^\sigma_{\sigma'}
$$
to first order in $\varpi$. We can now use \eqref{dspin2} to decompose $\Sigma_{S\mu\nu} = D^S(J_{\mu\nu})$ 
and take advantage of a known feature of the $D^S$ representation, namely that $D^S({\sf J}^i)$ 
are hermitian matrices while $D^S({\sf K}^i)$ are anti-hermitian, to find
\be\label{spindensexp}
  \Theta(p)^\sigma_{\sigma'} \simeq \frac{\delta^\sigma_{\sigma'}}{2S+1} + \frac{1}{2(2S+1)} 
  \varpi_*(p)^{\alpha\beta} \epsilon_{\alpha\beta\rho\nu} D^S({\sf J}^\rho)^\sigma_{\sigma'} 
  \hat t^\nu.
\ee
By plugging \eqref{spindensexp} into \eqref{meanspincov}, we get:
\begin{align}\label{vortspin}
 S^\mu(p) &= [p]^\mu_\kappa \frac{1}{2(2S+1)} \varpi_*(p)^{\alpha\beta} 
  \epsilon_{\alpha\beta\rho\nu} \tr \left( D^S({\sf J}^\rho) D^S({\sf J}^\kappa) \right) 
  \hat t^\nu \nonumber \\
  & = - \frac{1}{2(2S+1)}\frac{S(S+1)(2S+1)}{3} [p]^\mu_\kappa \varpi_*(p)^{\alpha\beta} 
  \epsilon_{\alpha\beta\rho\nu} g^{\rho \kappa} \hat t^\nu \nonumber \\ 
  & = - \frac{1}{2}\frac{S(S+1)}{3} [p]^\mu_\rho \varpi_*(p)_{\alpha\beta} 
  \epsilon^{\alpha\beta\rho\nu} \hat t_\nu = - \frac{1}{2m}\frac{S(S+1)}{3} \varpi_{\alpha\beta} 
  \epsilon^{\alpha\beta\mu\nu} p_\nu,
\end{align}
where, in the last equality, we have boosted the vector to the laboratory frame by using
the Eq.~\eqref{varpiboost}. 

For a fluid made of distinguishable particles, $\varpi$ becomes a local function \cite{becaspin}, 
so that the expression \eqref{vortspin} gives rise to the integral average:
\be\label{formula}
 S^\mu(p)= - \frac{1}{2m}\frac{S(S+1)}{3} \epsilon^{\mu\alpha\beta\nu} p_\nu
 \frac{\int d\Sigma_\lambda p^\lambda f(x,p) \varpi_{\alpha\beta}(x)}
 {\int d\Sigma_\lambda p^\lambda f(x,p)}
\ee
with $f(x,p)$ the distribution function. The latter is basically the same formula obtained in 
refs.~\cite{becaspin,becalisa}. 

\section{Spin density matrix and polarization transfer in two-body decays}\label{ptransfer}

Consider a massive particle, henceforth named ``Mother" and denoted by ``$M$", with spin $j$ and 
third component $m$ in its rest frame decaying into two particles, henceforth named ``Daughters" 
and denoted by ``$D_1$" and ``$D_2$". In the Mother rest frame, the magnitude of the momentum of 
the Daughters is fixed:
\be\label{pmagn}
  \p_{*}=\p_{*D}\equiv \frac{1}{2 m_M }\prod_{s,t=\pm}({m}_M +s\,m_{D_1}+t\,m_{D_2})^{1/2}
\ee
due to energy-momentum conservation, where $m_M$ is the mass of the Mother, and $m_{D_1}$, $m_{D_2}$ 
are the mass of the Daughters. As long as the decay is unobserved, the contribution of the decayed 
state to the quantum superposition reads, in the helicity basis \cite{Moussa,Tung,Chung},
\be\label{state}
   \ket{p_* j m \lambda_1 \lambda_2} \propto  T^j(\lambda_1,\lambda_2) \int \di \Omega_* \; 
   D^j(\varphi_*,\theta_*,0)^{m \, *}_{\lambda} \ket{{\bf p}_* \lambda_1\lambda_2}, 
\ee
where $\di \Omega_* = \sin\theta_* \di \theta_* \di \varphi_*$ is the infinitesimal solid angle 
corresponding to the momentum direction of Daughter $1$; $D^j(\varphi_*,\theta_*,0)^{m \, *}_{\lambda}$ 
is the complex conjugate of the rotation matrix element in the representation $j$ with components 
$m$ and $\lambda = \lambda_1 - \lambda_2$; and $T^j(\lambda_1,\lambda_2)$ are the reduced helicity 
transition amplitudes. 

Once a measurement of the momentum of either decay particle is made and a momentum ${\bf p}_*$
is found, according to quantum mechanics, the pure state gets reduced to a mixed one with density 
operator:
\be\label{rho1}
 \wrho({\bf p}_*) = \frac{T^j(\lambda_1,\lambda_2) T^j(\lambda'_1,\lambda'_2)^* 
   D^j(\varphi_*,\theta_*,0)^{m \, *}_{\lambda} D^j(\varphi_*,\theta_*,0)^{m}_{\lambda'}
   \ket{{\bf p}_* \lambda_1\lambda_2} \bra{{\bf p}_* \lambda'_1 \lambda'_2}}
   {|T^j(\lambda_1,\lambda_2)|^2 |D^j(\varphi_*,\theta_*,0)^{m}_{\lambda}|^2
   \braket{{\bf p}_* \lambda_1 \lambda_2}{{\bf p}_* \lambda_1 \lambda_2}}.
\ee
This is indeed the density operator in spin space for the two-Daughter system in the Mother's rest 
frame, for a given spin state $m$ of the Mother. However, the spin state of the Mother is 
itself a density operator of the sort \eqref{spindensf2} in the laboratory frame. This means 
that the decayed state is not simply \eqref{state}, rather the one with the density operator:
$$
  \sum_{m,n=-j}^j \Theta^m_{n} \ket{p_* jm \lambda_1 \lambda_2}\bra{p_* j n \lambda_1' \lambda_2'},
$$
hence the overall density operator of the Daughters reads explicitly:
\be\label{rho2}
 \wrho({\bf p}_*) \propto T^j(\lambda_1,\lambda_2) T^j(\lambda'_1,\lambda'_2)^* 
   D^j(\varphi_*,\theta_*,0)^{m \, *}_{\lambda} \Theta^m_n D^j(\varphi_*,\theta_*,0)^{n}_{\lambda'}
   \ket{{\bf p}_* \lambda_1\lambda_2} \bra{{\bf p}_* \lambda'_1 \lambda'_2}.
\ee

By using the density operator \eqref{rho2}, we can write down the normalized spin density matrix of 
the Daughters in the Mother rest frame as:
\be\label{spindensmat}
 \Theta_{D \, \lambda'_1 \lambda'_2}^{\; \; \lambda_1 \lambda_2} = 
 \frac{T^j(\lambda_1,\lambda_2) T^j(\lambda'_1,\lambda'_2)^* D^j(\varphi_*,\theta_*,0)^{m \, *}_{\lambda} 
 \Theta^m_n D^j(\varphi_*,\theta_*,0)^{n}_{\lambda'}}
 {|T^j(\lambda_1,\lambda_2)|^2 D^j(\varphi_*,\theta_*,0)^{m \, *}_{\lambda} 
 \Theta^m_n D^j(\varphi_*,\theta_*,0)^{n}_{\lambda}}, 
\ee
whence the mean spin vector of the Daughter 1 can be obtained from the Eq.~\eqref{meanspincov} as
\be\label{meanspind}
  S^\mu_1({\bf p_*}) = [p_*]^\mu_\nu D^{S1}({\sf J}^\nu)_{\lambda_1}^{\lambda'_1}
  \Theta_{D \, \lambda'_1 \lambda_2}^{\; \; \lambda_1 \lambda_2}.
\ee
Then, in general, the mean spin vector of the Daughters depends on the spin density matrix of
the Mother $\Theta$, which is a $(2j+1)\times(2j+1)$ hermitian matrix with trace 1 hence depending 
on $(2j+1)^2-1$ real parameters. For this reason, in principle, the mean spin vector of the Daughter
does not just depend on the mean spin vector of the Mother (3 real parameters), except for $j=1/2$
but it involves more variables, a well known fact in the literature \cite{Leader,Kim:1992az}. 
Nevertheless, in the case of relativistic nuclear collisions and QGP with small thermal vorticity, 
the spin density matrix of the primary Mothers can be approximated by the \eqref{spindensexp} 
which eventually entails that the mean spin vector of the Daughters depends only on the mean 
spin vector of the Mother, like in the spin-1/2 case, as we will see.

Let us implement the approximation \eqref{spindensexp} for $\Theta^m_n$ in the \eqref{spindensmat} 
to obtain its first order expansion in thermal vorticity. The first term in \eqref{spindensexp} 
is proportional to the identity and selects $m=n$ in \eqref{spindensmat}, then one is left with:
$$
  D^j(\varphi_*,\theta_*,0)^{m \, *}_{\lambda} D^j(\varphi_*,\theta_*,0)^{m}_{\lambda'}
  = \delta^\lambda_{\lambda'}
$$
due to the unitarity of the $D^j$'s. On the other hand, the second term in the \eqref{spindensexp} gives 
rise to the following product of three matrices:
$$
 D^j(\varphi_*,\theta_*,0)^{m \, *}_{\lambda} D^j({\sf J}^\rho)^m_n
 D^j(\varphi_*,\theta_*,0)^{n}_{\lambda'} =D^j(\varphi_*,\theta_*,0)^{-1 \, \lambda}_{m} 
 D^j({\sf J}^\rho)^m_n D^j(\varphi_*,\theta_*,0)^{n}_{\lambda'},
$$
which, according to a well known relation in group representation theory, equals
\be\label{rotred}
  {\sf R}(\varphi_*,\theta_*,0)^\rho_\tau  D^j({\sf J}^\tau)^\lambda_{\lambda'},
\ee
where the rotation ${\sf R}$ transforms the $z$-axis unit vector $\hat{\bf k}$ into the momentum ${\bf p}_*$
of the decayed particle. Altogether, we can rewrite the \eqref{spindensmat} as:
\be\label{spindensmat2}
 \Theta_{D \, \lambda'_1 \lambda_2}^{\; \; \lambda_1 \lambda_2} \simeq
 \frac{T^j(\lambda_1,\lambda_2) T^j(\lambda'_1,\lambda'_2)^*  
 \left[ \delta^{\lambda}_{\lambda'} + (1/2) \varpi_*(P)^{\alpha\beta} \epsilon_{\alpha\beta\rho\nu} 
 D^j({\sf J}^\tau)^\lambda_{\lambda'} {\sf R}(\varphi_*,\theta_*,0)^\rho_\tau {\hat t}^\nu \right]}
 {\sum_{\lambda_1,\lambda_2} |T^j(\lambda_1,\lambda_2)|^2},
\ee
where $P$ is the momentum of the Mother in the laboratory frame.

We are now in a position to work out \eqref{meanspind} by using the spin density 
matrix in the Eq.~\eqref{spindensmat2}. For strong and electromagnetic decays with parity conservation, it is readily found that the first term 
containing $\delta^\lambda_{\lambda'}$ does not give any contribution to the mean 
spin vector because $\lambda_2=\lambda_2'$ implies $\lambda_1 = \lambda'_1$ and one 
is left with a vanishing trace of $D^{S1}({\sf J}^\nu)$ in \eqref{meanspind}. Conversely, 
the second term in \eqref{spindensmat2} yields a finite non-vanishing result:
\begin{align}\label{meanspind2}
  S^\mu_1({\bf p_*}) & = \frac{1}{2} \varpi_*(P)^{\alpha\beta} \epsilon_{\alpha\beta\rho\nu} {\hat t}^\nu 
  \frac{T^j(\lambda_1,\lambda_2) T^j(\lambda'_1,\lambda_2)^*  
  [p_*]^\mu_\kappa D^{S_1}({\sf J}^\kappa)_{\lambda_1}^{\lambda'_1} D^j({\sf J}^\tau)^\lambda_{\lambda'} 
  {\sf R}(\varphi_*,\theta_*,0)^\rho_\tau}{\sum_{\lambda_1,\lambda_2} |T^j(\lambda_1,\lambda_2)|^2}
  \nonumber \\
  & = - \frac{3}{j(j+1)} S_{*M}(P)_\rho
  \frac{T^j(\lambda_1,\lambda_2) T^j(\lambda'_1,\lambda_2)^*  
  [p_*]^\mu_\kappa  D^{S_1}({\sf J}^\kappa)_{\lambda_1}^{\lambda'_1} D^j({\sf J}^\tau)^\lambda_{\lambda'} 
  {\sf R}(\varphi_*,\theta_*,0)^\rho_\tau}{\sum_{\lambda_1,\lambda_2} |T^j(\lambda_1,\lambda_2)|^2},
\end{align}
where we have used \eqref{vortspin} to express the formula in terms of the mean 
spin vector of the Mother in its rest frame $S_{*M}(P)$. In the next subsections, 
we will work out two specific relevant decays: 
$\Sigma^* \rightarrow \Lambda \pi$ and $\Sigma \rightarrow \Lambda \gamma$. 

\subsection{$\Sigma^* \rightarrow \Lambda \pi$}\label{Sigmas} 
	
In this case, $\lambda_2=0$, $j=3/2$ and $S_1 = 1/2$ in \eqref{meanspind2}, and $T(1/2,0)=T(-1/2,0)$
because of parity invariance \cite{becalisa}. Hence, there is only one independent helicity amplitude 
which cancels out in the \eqref{meanspind2} and we have:
\begin{align*}
   S^\mu_\Lambda({\bf p_*}) &= - \frac{3}{2j(j+1)} S_{*M}(P)_\rho
  [p_*]^\mu_\kappa  D^{1/2}({\sf J}^\kappa)_{\lambda_1}^{\lambda'_1} D^{3/2}
  ({\sf J}^\tau)^{\lambda_1}_{\lambda'_1} {\sf R}(\varphi_*,\theta_*,0)^\rho_\tau 
  \nonumber \\
  & =- \frac{3}{2j(j+1)} S_{*M}(P)_\rho [p_*]^\mu_\kappa  \tr ( D^{1/2}({\sf J}^\kappa) 
  D^{3/2}_{\rm RED}({\sf J}^\tau)) {\sf R}(\varphi_*,\theta_*,0)^\rho_\tau,
\end{align*}
where $D^{3/2}_{\rm RED}$ is the reduced $2\times2$ matrix formed with the elements labelled by the
indices $\lambda=\pm 1/2$.
Since
$$
  D^{3/2}_{\rm RED}({\sf J}^1)) = \sigma_1, \qquad D^{3/2}_{\rm RED}({\sf J}^2)) = \sigma_2,
  \qquad D^{3/2}_{\rm RED}({\sf J}^3)) = \frac{\sigma_3}{2},
$$
where the $\sigma$'s are the Pauli matrices, the mean spin vector becomes:
\begin{align*}
  S^\mu_\Lambda({\bf p_*}) &=- \frac{3}{2j(j+1)} S_{*M}(P)_\rho [p_*]^\mu_\kappa C_\tau \tr (\sigma_\kappa \sigma_\tau) 
  {\sf R}(\varphi_*,\theta_*,0)^\rho_\tau \nonumber \\
  & = \frac{3}{j(j+1)} S_{*M}(P)_\rho [p_*]^\mu_\kappa C_\tau g^{\kappa\tau}
  {\sf R}(\varphi_*,\theta_*,0)^\rho_\tau \nonumber \\
  &= \frac{3}{j(j+1)} S_{*M}(P)_\rho [p_*]^\mu_\tau C_\tau {\sf R}(\varphi_*,\theta_*,0)^{\rho \tau},
\end{align*}
with
$$
  C_\tau = (1/2,1/2,1/2,1/4) = \frac{1}{2} - \frac{1}{4} \delta_\tau^3. 
$$
Note that we set $C_0 = 1/2$ instead of the obvious $0$ as the multiplying matrix element 
${\sf R}(\varphi_*,\theta_*,0)^{\rho 0}$ always vanishes. 

In the helicity scheme, the matrix $[p_*]$ can be expanded according to \eqref{standard} and so, 
taking advantage of the orthogonality of rotations ${\sf R}$, we have
\begin{align*}
 S^\mu_\Lambda({\bf p_*}) &= \frac{3}{j(j+1)} S_{*M}(P)_\rho C_\tau {\sf R}(\varphi_*,\theta_*,0)^\mu_\nu 
 {\sf L}_z (\xi)^\nu_\tau {\sf R}^{-1}(\varphi_*,\theta_*,0)^{\tau \rho} \nonumber \\
 &= \frac{3}{2j(j+1)} \left[ L_{\bf \hat p_*}(\xi)^\mu_\rho S_{*M}(P)^\rho - \frac{1}{2} S_{*M}(P)_\rho
  {\sf R}(\varphi_*,\theta_*,0)^\mu_\nu {\sf L}_z (\xi)^\nu_3 {\sf R}(\varphi_*,\theta_*,0)^{\rho 3}
  \right],
\end{align*}
where $L_{\bf \hat p_*}(\xi)$ is the pure Lorentz boost transforming $\hat t$ into the direction
of ${\bf \hat p_*}$ in the Mother's rest frame. The Lorentz transformation can be expanded as well
by using the momentum of the Daughter:
\begin{align}\label{scalprod}
 {\sf R}(\varphi_*,\theta_*,0)^\mu_\nu {\sf L}_z (\xi)^\nu_3 {\sf R}(\varphi_*,\theta_*,0)^{\rho 3}
 ={}& {\sf R}(\varphi_*,\theta_*,0)^\mu_3 {\sf L}_z (\xi)^3_3 {\sf R}(\varphi_*,\theta_*,0)^{\rho 3}
 + {\sf R}(\varphi_*,\theta_*,0)^\mu_0 {\sf L}_z (\xi)^0_3 {\sf R}(\varphi_*,\theta_*,0)^{\rho 3}
 \nonumber \\
  ={} & - \cosh \xi \, {\bf \hat p_*}^\mu  {\bf \hat p_*}^\rho - \sinh \xi \, {\bf \hat p_*}^\rho \delta^\mu_0
  = - \frac{\varepsilon_*}{m_{\Lambda}} {\bf \hat p_*}^\mu {\bf \hat p_*}^\rho - 
  \frac{\p_*}{m_{\Lambda}} {\bf \hat p_*}^\rho \delta^\mu_0,
\end{align}
so that the spin vector becomes:
$$
  S^\mu_\Lambda({\bf p_*}) = \frac{3}{2j(j+1)} \left[ {\sf L}_{\bf \hat p_*}(\xi)^\mu_\rho S_{*M}(P)^\rho
  - \frac{\varepsilon_*}{2m_\Lambda} {\bf S}_{*M} \cdot {\bf \hat p_*} {\bf \hat p_*}^\mu 
  - \frac{\p_*}{2m_\Lambda} {\bf S}_{*M} \cdot {\bf \hat p_*} \delta^\mu_0 \right],
$$
where $m_\Lambda$ is the mass of the $\Lambda$. With the help of the known formulae of pure Lorentz boosts, 
we get:
\begin{align*}
  S^0_\Lambda({\bf p_*}) &= \frac{3}{2j(j+1)} \left[ \frac{1}{m_\Lambda}{\bf p}_* \cdot {\bf S}_{*M} 
  - \frac{1}{2m_\Lambda} {\bf S}_{*M} \cdot {\bf p_*} \right] = 
  \frac{3}{2j(j+1)} \frac{1}{2m_\Lambda} {\bf S}_{*M} \cdot {\bf p}_* , \\
  {\bf S}_\Lambda({\bf p_*}) &= \frac{3}{2j(j+1)} \left[ {\bf S}_{*M} + \frac{{\bf p_*} \cdot {\bf S}_{*M}}
  {(\varepsilon_*+ m_\Lambda)m_\Lambda} {\bf p_*} - 
  \frac{\varepsilon_*}{2m_\Lambda \p_*^2} {\bf S}_{*M} \cdot {\bf p_*} {\bf p_*} \right] 
  = \frac{3}{2j(j+1)} \left[ {\bf S}_{*M} + \frac{\varepsilon_*-2m_\Lambda}{2m_\Lambda \p_*^2}
  {\bf S}_{*M} \cdot {\bf p_*} {\bf p_*} \right]. 
\end{align*}
Finally, the spin vector is boosted to the $\Lambda$ rest frame and we use $j=3/2$ to obtain the 
spin three-vector as:
\be\label{lambdapol}
 {\bf S}_{0\Lambda}({\bf p_*}) = {\bf S}_\Lambda(p_*) - S^0_\Lambda(p_*) \frac{\bf p_*}{\varepsilon_*+m_\Lambda}
 = \frac{2}{5} \left[ {\bf S}_{*M} - \frac{1}{2} {\bf S}_{*M} \cdot {\bf \hat p_*} 
 {\bf \hat p_*} \right].
\ee
This result is in full agreement with the global mean spin vector of the $\Lambda$ from
polarized $\Sigma^*$ decay derived in ref.~\cite{becalisa}. Indeed, by setting ${\bf S}_{*M} = S_{*M} 
\hat {\bf k}$ without loss of generality and integrating over $\di \Omega_*/4 \pi$ to average out the angular dependence in 
\eqref{lambdapol}, we get:
$$
  \langle {\bf S}_{0\Lambda} \rangle = \frac{1}{3} {\bf S}_{*M}. 
$$
which is precisely the result found in ref.~\cite{becalisa}

\subsection{$\Sigma^0 \rightarrow \Lambda \gamma$}\label{Sigman}  

In this case, $j=1/2$, $S_1 = 1/2$, and $|\lambda_2|$ = 1 as the second particle 
is a photon. These numbers imply that $\lambda_2 =  2 \lambda_1$, $\lambda'_1=\lambda_1$ 
and $\lambda = \lambda_1 - \lambda_2 = - \lambda_1$~\cite{becalisa}. Furthermore, 
because of parity conservation, there is only one independent helicity
amplitude which cancels out and \eqref{meanspind2} becomes:
\be\label{spinslg1}
   S^\mu_\Lambda({\bf p_*}) = - \frac{3}{2j(j+1)} S_{*M}(P)_\rho
 [p_*]^\mu_\kappa  D^{1/2}({\sf J}^\kappa)^{\lambda_1}_{\lambda_1} D^{1/2}
  ({\sf J}^\tau)_{-\lambda_1}^{-\lambda_1} {\sf R}(\varphi_*,\theta_*,0)^\rho_\tau. 
\ee
The combination 
$$
  D^{1/2}({\sf J}^\kappa)_{-\lambda_1}^{-\lambda_1} D^{1/2}
  ({\sf J}^\tau)^{\lambda_1}_{\lambda_1} = C_\tau g^{\kappa\tau}
$$
vanishes except for $\kappa=\tau=3$ with
$$
  C_\tau = (0,0,0,1/2) = \frac{1}{2} \delta^3_\tau,
$$
then we can rewrite \eqref{spinslg1} as:
\begin{align*}
  S^\mu_\Lambda({\bf p_*}) &= -\frac{3}{2j(j+1)} S_{*M}(P)_\rho [p_*]^\mu_\kappa C_\tau g^{\kappa\tau}
  {\sf R}(\varphi_*,\theta_*,0)^\rho_\tau =-\frac{3}{4j(j+1)} S_{*M}(P)_\rho [p_*]^\mu_\tau \delta^3_\tau 
  {\sf R}(\varphi_*,\theta_*,0)^{\rho \tau} \nonumber \\ 
  &=-\frac{3}{4j(j+1)} S_{*M}(P)_\rho [p_*]^\mu_3 {\sf R}(\varphi_*,\theta_*,0)^{\rho 3}
  =-\frac{3}{4j(j+1)} S_{*M}(P)_\rho {\sf R}(\varphi_*,\theta_*,0)^\mu_\nu
   {\sf L}_3(\xi)^\nu_3 {\sf R}(\varphi_*,\theta_*,0)^{\rho 3}.
\end{align*}
By plugging the \eqref{scalprod}, we find
$$
  S^\mu_\Lambda({\bf p_*}) =  - \frac{3}{4j(j+1)} {\bf S}_{*M} \cdot {\bf \hat p_*}  
  \left(\frac{\varepsilon_*}{m_{\Lambda}}  {\bf \hat p_*}^\mu + \frac{\p_*}{m_{\Lambda}} \delta^\mu_0 \right),
$$
which yields the three-component spin vector:
\be\label{lambdapol2}
 {\bf S}_{0\Lambda}({\bf p_*}) = {\bf S}_\Lambda(p_*) - S^0_\Lambda(p_*) \frac{\bf p_*}{\varepsilon_*+m_\Lambda}
 = - {\bf S}_{*M} \cdot {\bf \hat p_*} \, {\bf \hat p_*}
\ee
after boosting back to the $\Lambda$ rest frame. Similar to the $\Sigma^*$ case, this equation is again in full agreement with the {\em average} 
spin vector of the $\Lambda$ from polarized $\Sigma^0$ decay~\cite{becalisa}.

\section{Momentum average and longitudinal polarization}
\label{sec:momentum}

So far, we have derived the formulae for the polarization transfer in two-body decays in terms of the mean 
spin vectors of particles in their rest frames  - what experiments can actually measure - for some given momentum
of the $\Lambda$ in the Mother's rest frame. However, we are more interested in the polarization vector 
inherited by the $\Lambda$ as a function of the $\Lambda$ momentum ${\bf p}$ in the laboratory frame.
Besides, Mother particles have a momentum distribution and we have to fold our result therewith in order 
to compare with the experimental results. 

Let ${\bf P}$ be the momentum of the Mother in the laboratory frame and $n({\bf P})$ its un-normalized momentum
spectrum, such that ${\int \di^3 {\rm P} \; n({\bf P})}$ yields the total number of the Mother. One would 
then define the mean spin vector of the $\Lambda$ for a specific decay as:
$$
  \langle {\bf S}_{0\Lambda}({\bf p}) \rangle = 
  \frac{\int \di^3 {\rm P} \; n({\bf P}) \, {\bf S}_{0\Lambda}({\bf p_*})}{\int \di^3 {\rm P} \; n({\bf P})},
$$
where ${\bf S}_{0\Lambda}({\bf p_*})$ is given by either \eqref{lambdapol} or \eqref{lambdapol2} and ${\bf p_*}$
is a function of ${\bf p}$ and ${\bf P}$ because it is related to ${\bf p}$ by a Lorentz transformation. 
Since the magnitude of ${\bf p}_*$ is fixed, the three components of ${\bf P}$ are not completely independent 
for a given ${\bf p}$. This can be seen from the Lorentz transformation relation:
$$
 \varepsilon_* = \frac{\varepsilon_M}{m_M} \varepsilon - \frac{1}{m_M} {\bf P} \cdot {\bf p} = \sqrt{\p^2_{*D} + m^2_\Lambda},
$$
where $\varepsilon = \sqrt{\p^2 + m_\Lambda^2}$ and $\p_{*D}$ is given by the \eqref{pmagn}. To implement this constraint, one should multiply both the
numerator and denominator of the above equation by a $\delta(\p_*-\p_{*D})$ so that:
$$
  \langle {\bf S}_{0\Lambda}({\bf p}) \rangle = 
  \frac{\int \di^3 {\rm P} \; n({\bf P}) \, {\bf S}_{0\Lambda}({\bf p}_*) \delta(\p_* - \p_{*D})}
  {\int \di^3 {\rm P} \; n({\bf P}) \delta(\p_* - \p_{*D})}.
$$
Now, it is convenient to change the integration variable from ${\bf P}$ to ${\bf p}_*$ to take advantage
of the straightforward $\delta$ integration and this can indeed be done by using Lorentz boosts
(see Appendix~\ref{app:lorentz}):
\be\label{momrel}
 {\bf P} = 2 m_M \frac{(\varepsilon_* + \varepsilon)({\bf p} - {\bf p_*})}
 {(\varepsilon_* + \varepsilon)^2-({\bf p} - {\bf p_*})^2} = 
  m_M \frac{(\varepsilon_* + \varepsilon)({\bf p} - {\bf p_*})}
 {m_\Lambda^2 + \varepsilon \varepsilon_* +{\bf p}\cdot{\bf p_*}}
 \implies 
 \hat{\bf P} = \frac{{\bf p} - {\bf p_*}}{\|{\bf p} - {\bf p_*}\|},
\ee
After this change of variable, the integration in $\di \p_*$ is trivial and we are left with the 
solid angle integration:
\be\label{decayspin}
  \langle {\bf S}_{0\Lambda}({\bf p}) \rangle = 
  \frac{\int \di^3 \p_* \; n({\bf P})  \left| \frac{\partial {\bf P}}{\partial{\bf p}_*} \right|\, 
  {\bf S}_{0\Lambda}({\bf p}_*) \delta(\p_* - \p_{*D})}
  {\int \di^3 \p_* \; n({\bf P})  \left| \frac{\partial \bf P}{\partial {\bf p}_*} \right| \, 
  \delta(\p_* - \p_{*D})} =  
  \frac{\int \di \Omega_* \; n({\bf P}) \left| \frac{\partial {\bf P}}{\partial{\bf p}_*} \right|\, 
  {\bf S}_{0\Lambda}({\bf p}_*)}
  {\int \di \Omega_* n({\bf P})  \left| \frac{\partial \bf P}{\partial {\bf p}_*} \right| \,},
\ee
where the absolute value of the determinant of the 
Jacobian reads (see Appendix~\ref{app:lorentz}):
\be\label{jacob}
\left| \frac{\partial {\bf P}}{\partial{\bf p}_*} \right|= \frac{m_M^3(\varepsilon_* + \varepsilon)^2 \left[ (\varepsilon_* + \varepsilon)^2 - (\varepsilon \varepsilon_* + {\bf p}\cdot {\bf p}_*+m_\Lambda^2) \right]}{\varepsilon_*(\varepsilon \varepsilon_* + {\bf p}\cdot {\bf p}_*+m_\Lambda^2)^3}.
\ee

The ${\bf S}_{0\Lambda}({\bf p}_*)$ in eqs. \eqref{lambdapol} and \eqref{lambdapol2} are now 
to be used in the \eqref{decayspin} to obtain the expressions of the mean spin of the $\Lambda$ 
inherited in the decays of $\Sigma^*$ and $\Sigma^0$ produced in the nuclear collision:
\begin{align}
  \langle {\bf S}_{0\Lambda}({\bf p}) \rangle= & 
  \frac{\int \di \Omega_* \; n({\bf P})  \left| \frac{\partial {\bf P}}{\partial{\bf p}_*} \right| \, 
  \frac{2}{5} \left[ {\bf S}_{*M} - \frac{1}{2} {\bf S}_{*M} \cdot {\bf \hat p_*} 
 {\bf \hat p_*} \right]}
  {\int \di \Omega_* n({\bf P})  \left| \frac{\partial \bf P}{\partial {\bf p}_*} \right| \,}
  \qquad  \qquad&\Sigma^* \to \Lambda  \pi,
  \nonumber \\
 \langle {\bf S}_{0\Lambda}({\bf p}) \rangle =& - 
  \frac{\int \di \Omega_* \; n({\bf P})  \left| \frac{\partial \bf P}{\partial {\bf p}_*} \right| \, 
   {\bf S}_{*M} \cdot {\bf \hat p_*} {\bf \hat p_*}}
  {\int \di \Omega_* n({\bf P})  \left| \frac{\partial {\bf P}}{\partial{\bf p}_*} \right| \,}
  \qquad\qquad& \Sigma^0 \to \Lambda  \gamma,
  \label{decayspin2}
  \end{align}
As it can be seen, the contribution to the $\Lambda$ polarization is proportional to the mean spin 
vector of the Mother, just like the global polarization, but in a more complicated fashion than simple 
vector collinearity.

Because of the symmetries of the system of colliding nuclei in peripheral collisions (total 
reflection and rotation around the total angular momentum axis), the components of the polarization 
vector should exhibit a symmetry pattern in momentum space (see also discussion in ref.~\cite{becakarp}).
Particularly, they can be expanded in Fourier series as a function of the momentum azimuthal angle
and, because of those symmetries, the leading harmonics are~\cite{becavort,becakarp,xia,xiatalk}
\begin{align}\label{harmonics}
 S_{*Mx} &\simeq \frac{2j(j+1)}{3} \left[ h_1({\rm P}_T,Y) \sin \varphi_M
 + h_2({\rm P}_T,Y) \sin 2 \varphi_M \right] ,  \nonumber \\
 S_{*My} &\simeq \frac{2j(j+1)}{3} \left[ g_0(P_T,Y) + g_1({\rm P}_T,Y) \cos \varphi_M 
 + g_2({\rm P}_T,Y) \cos 2 \varphi_M \right], \nonumber \\
 S_{*Mz} &\simeq \frac{2j(j+1)}{3} f_2({\rm P}_T,Y) \sin 2 \varphi_M,
\end{align}
where $Y$ is the rapidity of the Mother. The coefficient ${2j(j+1)/3}$ in front is meant to
remove from the functions $f,g,h$ their trivial dependence on the spin, according to the
eq.~\eqref{vortspin}. The aforementioned symmetries imply that $h_1$ and
$g_1$ are odd functions of $Y$ whereas $g_0, f_2, g_2, h_2$ are even. The hydrodynamic model 
supplied with the usual initial conditions \cite{becavort,xiatalk} predicts the magnitude of 
the functions $f,g,h$ and particularly their signs to be:
$$
 g_0 < 0, \qquad h_2 < 0, \qquad g_2 > 0, \qquad f_2 < 0 ,
$$
in a right-handed reference frame with $x$-axis on the reaction plane and $y$-axis in the
direction opposite to the total angular momentum. 
We can decompose the momenta in spherical coordinates and use the rightmost equality in Eq.~\eqref{momrel}, 
\begin{align}
    {\bf P} ={}& {\rm P}_T \cos \varphi_M \hat{\bf i} + {\rm P}_T \sin \varphi_M \hat{\bf j} +
    {\rm P}_z \hat{\bf k},  \nonumber \\
    {\bf p}={}& \p_T \cos \varphi \hat{\bf i} + \p_T \sin \varphi \hat{\bf j} + \p_z \hat{\bf k},\nonumber \\
    {\bf p_*}={}& \p_* \sin \theta_* \cos \varphi_* \hat{\bf i} + \p_* \sin \theta_* \sin \varphi_* \hat{\bf j} 
    + \p_* \cos \theta_* \hat{\bf k},
\end{align}
where the trigonometric functions can be expressed in terms of the momenta of the $\Lambda$ as:
\bea\label{phimother}
  \sin 2 \varphi_M  & = & \frac{ \p_*^2 \sin^2\theta_* \sin 2 \varphi_* + \p_T^2 
  \sin 2\varphi - 2 \p_* \p_T \sin \theta_* \sin (\varphi_* +\varphi)}
  {\p_*^2 \sin^2\theta_*+\p_T^2 - 2\p_* \p_T \sin \theta_* \cos (\varphi_* -\varphi)} =
  {\cal A}(\theta_*,\psi)\sin2\varphi +{\cal B}(\theta_*,\psi)\cos2\varphi, \nonumber\\
  \cos 2 \varphi_M  &=& \frac{\p_*^2 \sin^2\theta_* \cos 2 \varphi_* + \p_T^2 \cos 2\varphi - 
  2 \p_* \p_T \sin \theta_* \cos(\varphi_*+\varphi)}{\p_*^2 \sin^2\theta_*+\p_T^2 - 2\p_* \p_T 
  \sin \theta_* \cos (\varphi_* -\varphi)}
  ={\cal A}(\theta_*,\psi)\cos 2\varphi - {\cal B}(\theta_*,\psi)\sin2\varphi,  \nonumber \\
   \sin \varphi_M  & = & \frac{\p_T 
 	\sin \varphi - \p_* \sin\theta_* \sin \varphi_*}
 {\sqrt{\p_*^2 \sin^2\theta_*+\p_T^2 - 2\p_* \p_T \sin \theta_* \cos (\varphi_* -\varphi)}} =
 {\cal C}(\theta_*,\psi)\sin \varphi +{\cal D}(\theta_*,\psi)\cos \varphi, \nonumber\\
 \cos \varphi_M  &=& \frac{\p_T 
 \cos \varphi - \p_* \sin\theta_* \cos \varphi_*}{\sqrt{\p_*^2 \sin^2\theta_*+\p_T^2 - 2\p_* 
  \p_T \sin \theta_* \cos (\varphi_* -\varphi)}}
= {\cal C}(\theta_*,\psi)\cos \varphi - {\cal D}(\theta_*,\psi)\sin \varphi
\eea
with the variable $\psi = \varphi_* - \varphi $ and the auxiliary functions: 
\bea\label{coeff}
 {\cal A}(\theta_*,\psi)&=&{\p_*^2\sin^2\theta_* \cos 2 \psi - 2 \p_* \p_T \sin\theta_* \cos\psi +\p_T^2 
  \over \p_*^2 \sin^2\theta_*+\p_T^2 - 2\p_* \p_T \sin \theta_* \cos \psi},\nonumber\\
 {\cal B}(\theta_*,\psi)&=&{\p_*^2 \sin^2\theta_* \sin 2\psi - 2\p_* \p_T \sin\theta_* \sin\psi 
 \over \p_*^2 \sin^2\theta_*+\p_T^2 - 2\p_* \p_T \sin \theta_* \cos \psi}, \nonumber \\
{\cal C}(\theta_*,\psi)&=&{\p_T - \p_* \sin\theta_* \cos \psi 
	\over \sqrt{\p_*^2 \sin^2\theta_*+\p_T^2 - 2\p_* \p_T \sin \theta_* \cos \psi}},\nonumber\\
{\cal D}(\theta_*,\psi)&=&{- \p_* \sin\theta_* \sin \psi  
	\over \sqrt{\p_*^2 \sin^2\theta_*+\p_T^2 - 2\p_* \p_T \sin \theta_* \cos \psi}}.
\eea
The integration over the solid angle $\di \Omega_*$ in \eqref{decayspin2} can be replaced by 
an integration over $\psi$:
$$
 \int \di \Omega_* = \int_0^\pi \di \theta_* \sin \theta_* \int_{0}^{2\pi} \di \varphi_*
 = \int_0^\pi \di \theta_* \sin \theta_* \int^{2\pi-\varphi}_{-\varphi} \di \psi =
 \int_0^\pi \di \theta_* \sin \theta_* \int_{-\pi}^{\pi} \di \psi, 
$$
where the last equality is owing to the $2\pi$-periodic in $\psi$ of all integrand functions. 

We now specialize to the case of midrapidity $\Lambda$'s, that is, with $\p_z = {\bf p} \cdot \hat{\bf k} = 0$.
The spectrum function $n({\bf P})$ depends on the specific model of the collision, but it must be {\em even} 
in $\cos \theta_M$ because of the symmetries of the colliding system, and, to the approximation
of neglecting elliptic flow, isotropic in $\varphi_M$, that is depending only on ${\rm P}_T$ to 
a very good approximation. From \eqref{momrel} we have:
\be\label{ptran}
 {\rm P}_T = m_M \frac{(\varepsilon_* + \varepsilon) 
  \sqrt{\p_*^2 \sin^2\theta_* + \p_T^2 - 2\p_T \p_* \sin\theta_* \cos\psi}} 
  {m_\Lambda^2 + \varepsilon \varepsilon_* + \p_T \p_{*} \sin\theta_* \cos\psi},
\ee
which implies that the Mother spectrum function is even in $\cos \theta_*$ and even in $\psi$.
For mid-rapidity $\Lambda$ with $\p_z=0$, it ensues from \eqref{momrel} that:
\be\label{plong}
  \hat{\bf P} \cdot \hat{\bf k} = \cos \theta_M =  
  - \frac{\p_*\cos \theta_*}{\|{\bf p} - {\bf p_*}\|} =  - \frac{\p_*\cos \theta_*}{\sqrt{  
  \p_*^2+\p_T^2-2\p_T \p_* \sin\theta_* \cos\psi}},
\ee  
so the Mother spectrum function must be an even function of $\cos \theta_*$ as well.
Moreover, the rapidity $Y$ is an odd function of $\cos \theta_*$ and an even function of 
$\psi$ because:
$$
  M_T \sinh Y = \sqrt{{\rm P}_T^2 + m_M^2} \sinh Y = {\bf P} \cdot \hat{\bf k} 
  = {\rm P}_T \tan \theta_M
$$
in view of the equations \eqref{ptran} and \eqref{plong}. Similarly, it can be shown that 
the determinant of the Jacobian \eqref{jacob} is also even in $\psi$ and $\cos \theta_*$.

We now focus on the polarization along the beam, i.e., $\langle {\bf S}_{0\Lambda}({\bf p}) \rangle 
\cdot \hat{\bf k}$. Formulae \eqref{decayspin2} can then be written in a compact form as:
$$
  \langle {\bf S}_{0\Lambda}({\bf p}) \rangle \cdot \hat{\bf k} =  
  \frac{\int \di \Omega_* \; n({\bf P})  \left| \frac{\partial {\bf P}}{\partial{\bf p}_*} \right| \, 
  \left[ A~S_{*M z} + B~( {\bf S}_{*M} \cdot {\bf \hat p_*} \cos \theta_*) \right]}
  {\int \di \Omega_* n({\bf P})  \left| \frac{\partial {\bf P}}{\partial{\bf p}_*} \right| \,} 
$$
with $A$ and $B$ depending on the specific decay, namely:
\be\label{ab}
  A = \frac{2}{5} \qquad B = -\frac{1}{5}  \qquad {\rm for}\; \Sigma^*, \qquad \qquad \qquad
  A = 0           \qquad B = - 1  \qquad {\rm for}\; \Sigma^0. \qquad \qquad \qquad
\ee
We thus have:
\be\label{longspin1}
  \langle {\bf S}_{0\Lambda}({\bf p}) \rangle \cdot \hat{\bf k} =
  \frac{\int \di \Omega_* \; n({\bf P})  \left| \frac{\partial {\bf P}}{\partial{\bf p}_*} \right| \, 
 \left[ S_{*M z} \left( A + B \cos^2 \theta_* \right) + \frac{B}{2} \left( S_{*M x} \cos \varphi_* \sin 2 \theta_*
  + S_{*M y} \sin \varphi_* \sin 2 \theta_* \right) \right] }
  {\int \di \Omega_* n({\bf P})  \left| \frac{\partial {\bf P}}{\partial{\bf p}_*} \right| \,}.
\ee
To work out Eq.~\eqref{longspin1} one has to use \eqref{harmonics}, then \eqref{phimother} and 
\eqref{coeff}. The symmetry features in $\theta_*$ and $\psi$ simplify the computation because,
as the spectrum and the determinant are even functions of $\cos \theta_*$, and the coefficients 
${\cal A},{\cal B},{\cal C},{\cal D}$ in \eqref{coeff} as well, only few terms survive the integration 
in $\theta_*$. Particularly, the terms proportional to functions which are rapidity-even such as 
$g_0, h_2, g_2$ vanish as they multiply the odd function $\sin 2 \theta_*$. Besides, since:
\begin{align}\label{psiparity}
    {\cal A}(\theta_*,-\psi) &= {\cal A}(\theta_*,\psi) \qquad \qquad {\cal B}(\theta_*,-\psi) 
    = -{\cal B}(\theta_*,\psi) \\
     {\cal C}(\theta_*,-\psi) &= {\cal C}(\theta_*,\psi) \qquad \qquad \, {\cal D}(\theta_*,-\psi) 
     = -{\cal D}(\theta_*,\psi)  
\end{align}
all combinations involving odd functions in $\psi$ will vanish after integration. We are thus
left with the following function as integrand:
\be\label{integrand}
 f_2 ({\rm P}_T,Y) \left( A + B \cos^2 \theta_* \right) {\cal A}(\theta_*,\psi) \sin 2 \varphi
 + \frac{B}{4} \left[ h_1({\rm P}_T,Y) + g_1({\rm P}_T,Y) \right] 
 \left[\mathcal{C}(\theta_*,\psi) \cos\psi - \mathcal{D}(\theta_*,\psi)\sin\psi\right]  \sin 2\theta_* \sin 2\varphi,
\ee
which, as expected, is altogether proportional to $\sin 2 \varphi$, like for the primary hyperons.

At sufficiently high energy, because of approximate longitudinal boost invariance, we expect 
all of the functions $g,h,f$ in Eq.~\eqref{harmonics} to be very weakly dependent on rapidity.
As a consequence, since $h_1$ and $g_1$ are rapidity-odd, they should vanish at $Y=0$, hence 
negligible compared to $f_2$ and $g_2$. Therefore, we can approximate the longitudinal component of 
the transferred spin as:
\begin{align}\label{longspin2}
  \langle {\bf S}_{0\Lambda}({\bf p}_T) \rangle \cdot \hat{\bf k} & \simeq \sin 2 \varphi
  \, \frac{2j(j+1)}{3}
  \frac{\int_0^\pi \di \theta_* \sin \theta_* \int_{-\pi}^\pi \di \psi \; n({\bf P})  
  \left| \frac{\partial {\bf P}}{\partial{\bf p}_*} \right| \, 
  f_2 ({\rm P}_T) \left( A + B \cos^2 \theta_* \right) {\cal A}(\theta_*,\psi)}
  {\int_0^\pi \di \theta_* \sin \theta_* \int_{-\pi}^\pi \di \psi \;   n({\bf P})  
 \left| \frac{\partial {\bf P}}{\partial{\bf p}_*} \right| \,} \nonumber \\
  &\equiv \sin 2 \varphi \, \frac{2j(j+1)}{3} f_2 ({\p}_T)_{\rm M \to \Lambda}.
\end{align}
Putting all contributions together, including primaries:
\be\label{total}
  \langle \langle {\bf S}_{0\Lambda}({\bf p}_T) \rangle \cdot \hat{\bf k}\rangle =
  {1\over2}F_2^{\rm Tot}(\p_T)\sin 2 \varphi \equiv \frac12
   \left[ F_2 (\p_T) + F_2 ({\p}_T)_{\rm \Sigma^* \to \Lambda}
   + F_2 ({\p}_T)_{\rm \Sigma^0 \to \Lambda} \right] \sin 2 \varphi,
\ee
with 
$$
F_2 (\p_T) = X_p  f_2 (\p_T), \qquad \qquad F_2 ({\p}_T)_{\rm \Sigma^* \to \Lambda}=5 X_{\Sigma^*}  
f_2 ({\p}_T)_{\rm \Sigma^* \to \Lambda}, \qquad \qquad F_2 ({\p}_T)_{\rm \Sigma^0 \to \Lambda} =
X_{\Sigma^0} f_2 ({\p}_T)_{\rm \Sigma^0 \to \Lambda} ,
$$
where $X$'s are the $\Lambda$ number fractions from the different contributing channels:
primary $X_p$ and secondary $X_{\Sigma^*}$, $X_{\Sigma^0}$.

Likewise, it is possible to calculate the component of the overall mean spin vector along 
different directions, notably the total angular momentum direction, conventionally $-y$. 
Unlike the longitudinal component, it involves at least two of the functions in Eq.~\eqref{harmonics}, 
that is $g_0$ and $g_2$ (see Appendix.~\ref{App:Tpolarization}); this makes numerical computation 
more difficult than longitudinal component and we do not carry it out here, leaving it to 
future work.

\section{Numerical computation and results}\label{results}

We are now in a position to evaluate the contribution of the secondary $\Lambda$'s to the whole 
longitudinal polarization by using \eqref{total}, under the assumption of no re-interaction which
means no loss of mean polarization due to final state elastic rescattering of hadrons. The 
fractions of primary and secondary $\Lambda$'s can be estimated by means of the statistical hadronization 
model. At the hadronization temperature $T=164$ MeV and baryon chemical potential of 30 MeV 
for $\snn =200$ GeV \cite{SHM}, these fractions turn out to be $X_p= 0.243$, $X_{\Sigma^*} = 0.359$ 
and $X_{\Sigma^0}=0.275$ with a primary fraction of $\Sigma^*$ of virtually 100\% and of $\Sigma^0$ 
around 60\%, which has been used as a further factor in our computations. Furthermore, at this 
temperature the quantum statistics effects are negligible for all of these particles and the 
distinguishable particle approach used in this paper is an excellent approximation.

\begin{figure}[!htb]
	\begin{center}
		\includegraphics[width=8cm]{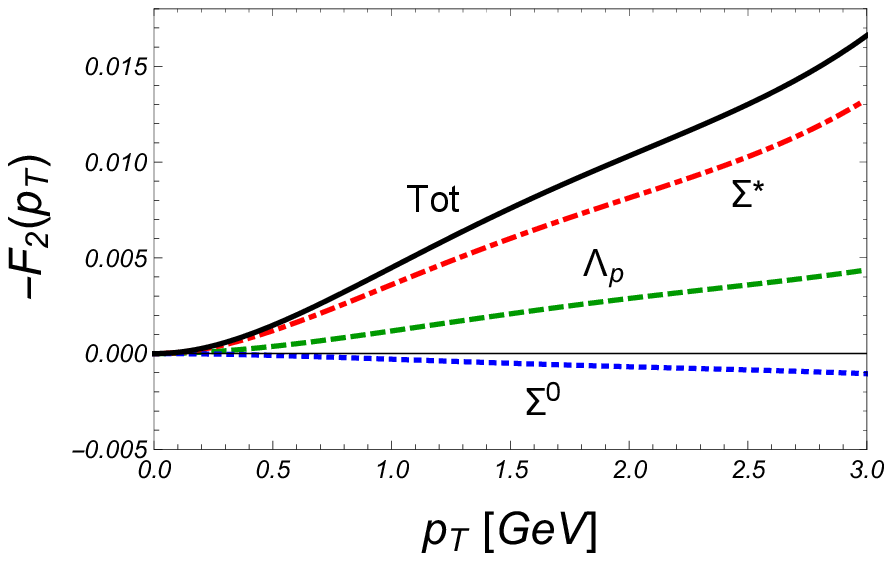}
		\includegraphics[width=8cm]{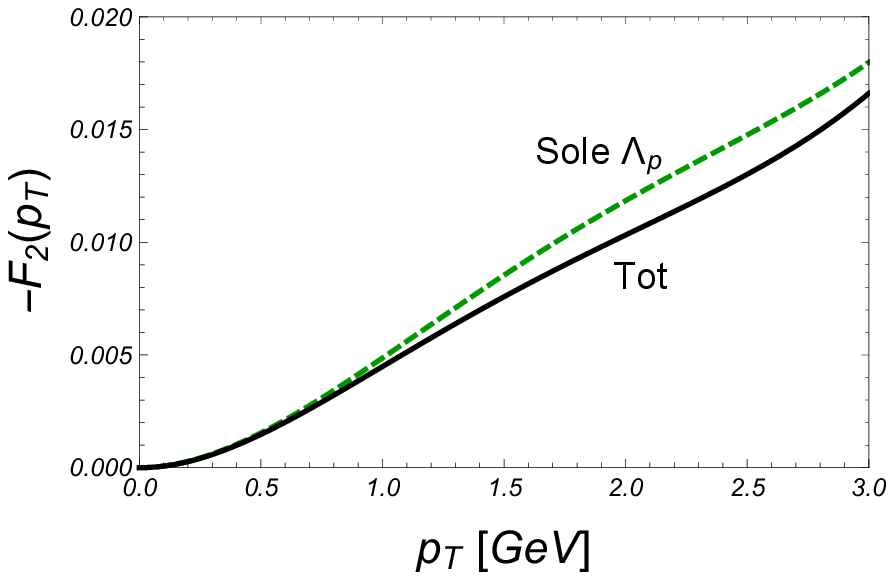}
		\caption{(color online) Left panel: longitudinal polarization coefficients $F_2(\p_T)$ of the $\Lambda$.
		Primary and secondary components, weighted with the production fractions are shown together with the 
		resulting sum (solid line). Right panel: comparison between the total polarization coefficient 
		$F_2^{\rm Tot}(\p_T)$ of the $\Lambda$ and the one $f_2(\p_T)$ of only primary $\Lambda$'s~\cite{becakarp}. }
		\label{f2}
	\end{center}
\end{figure}

\begin{figure}[!htb]
	\begin{center}
	\includegraphics[width=8cm]{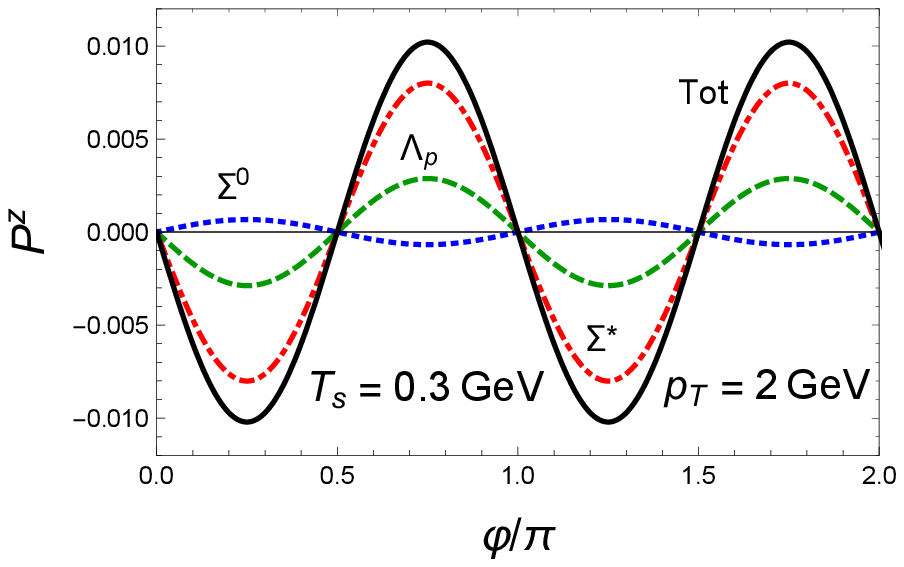}
	\includegraphics[width=8cm]{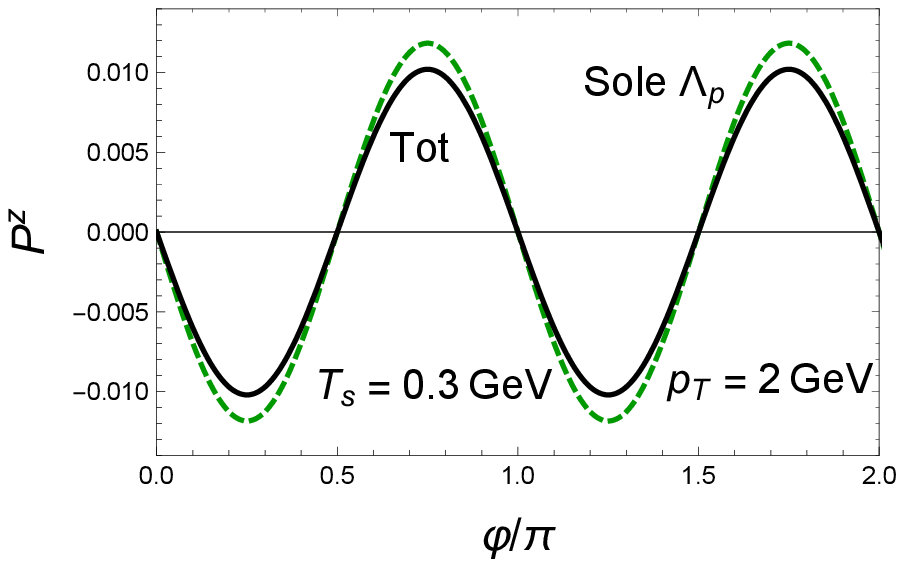}
		\caption{(color online) Left panel: the azimuthal angle dependence of the longitudinal polarization 
		$P^z = \langle {\bf S}_{0\Lambda} \rangle \cdot \hat{\bf k}/S = 2 \langle {\bf S}_{0\Lambda} \rangle \cdot \hat{\bf k}$ of the $\Lambda$. Primary and secondary components, weighted with the production 
		fractions are shown
		together with the resulting sum (solid line) at fixed transverse momentum $\p_T=2~{\rm GeV}$ and 
		slope parameter $T_s=0.3~{\rm GeV}$. Right panel: comparison between the total polarization of 
		the $\Lambda$ including secondary contributions with that of only primary $\Lambda$'s ~\cite{becakarp}.}
		\label{tot1}
	\end{center}
\end{figure}

The other ingredients in \eqref{total} are the two-body transfer functions $F_2$'s  which require, 
according to \eqref{longspin2}, the primary $f_2$'s of the Mother $\Sigma^*$ and $\Sigma^0$ as well as their 
momentum spectrum $n({\bf P})$. The primary $f_2$ functions depend mostly on the thermal vorticity and weakly on 
the mass of the hadrons through the Fermi-Dirac distribution function (see \eqref{distribution}), hence 
it is a very good approximation to take the same function for all involved hyperons, that is, $\Lambda$,
$\Sigma^0$ and $\Sigma^*$. A fit to the $f_2(\p_T)$ function of the primary $\Lambda$'s obtained in 
Ref.~\cite{becakarp} yields:
$$
f_2(\p_T)=-7.71 \times 10^{-3} \p_T^2+ 3.32 \times 10^{-3} \p_T^3 - 4.71 \times 10^{-4} 
\p_T^4,
$$
with $\p_T$'s units ${\rm GeV}$. The functions $F_2(\p_T)$ for primary and secondary decay 
contributions are shown in Fig.~\ref{f2}, and the associated polarizations are shown 
in Fig.~\ref{tot1} where we choose $\p_T=2~{\rm GeV}$ as an example.

As far as the momentum spectrum is concerned, because of its appearance in both numerator and 
denominator of \eqref{longspin2}, it is plausible that the dependence on its shape is very mild. For the
purpose of an approximate calculation, we have assumed a spectrum of the following form:
\be\label{distribution}
 n({\bf P}) \propto \frac{1}{\cosh y_M} \e^{-M_T/T_s} = \frac{m_M}{\varepsilon_M} \e^{-M_T/T_s},
\ee
where $y_M$ is the rapidity of the Mother, the transverse mass $M_T = \sqrt{{\rm P}_T^2 + m_M^2}$ and
$T_s$ is a phenomenological parameter describing the slope of the transverse momentum spectrum. We
have checked that the final results are very weakly dependent on $T_s$ within a realistic range
between $0.2$ and $0.8$ ${\rm GeV}$.

The results are shown in Figures~\ref{f2} and \ref{tot1}. It can be seen that the weighted sum of
all the $f_2$ components gives rise to almost the same result as for the primary $\Lambda$ alone.
This is somewhat surprising because the signs of the components originated from $\Sigma^*$ and 
$\Sigma^0$ differ in sign and magnitude. In fact, the quasi-coincidence observed on the right panels 
of the figures is seemingly an accidental result of the combinations of magnitudes of the various
terms and the fractions $X$ in Eq.~\eqref{total}. There are missing contributions from secondary
$\Sigma^0$ and more resonances decaying into $\Lambda$, but their weight in $\Lambda$ production is 
altogether limited compared to the computed one and definitely not able to flip the sign of $f_2$ 
in figure \ref{f2}.

\section{Summary and conclusions}\label{conclusions}

In summary, we have studied the contribution to the polarization of $\Lambda$ hyperons from the
decays of $\Sigma^*$ and $\Sigma^0$, which are the main sources of secondary $\Lambda$'s in nuclear 
collisions. Particularly, we have studied the corrections to the primary pattern of the longitudinal component 
(i.e. along the beam line) of the spin vector and found out that, provided that polarization 
is weakly dependent on rapidity, its cumulative effect is small compared to almost all previous 
calculations with only primary contribution. This is apparently the result of an accidental 
combination of primary and secondary fractions, and of the spins of $\Sigma^*$ and $\Sigma^0$. 
Since other contributions to $\Lambda$ production are almost negligible, we conclude that
hadronic decays cannot account for the observed discrepancy between the experimental results and 
the predictions of the thermodynamic-hydrodynamic model on the longitudinal polarization of the 
$\Lambda$ at $\snn = 200$ GeV \cite{niida}.

\section*{Acknowledgments}

We acknowledge very interesting discussions with many participants in the {\em Chirality, vorticity
and magnetic fields in heavy ion collisions} conference in Beijing, April 8-12 2019, and especially
L. Csernai, X. G. Huang, M. Lisa, S. Voloshin, Q. Wang. Therein, we got to know that X.G. Huang, 
H.Z. Huang, H. Li and X.L. Xia were working on the same topic \cite{huili}. G.C. is supported 
by the NSFC grant 11805290. E.S.\ is supported by the Deutsche Forschungsgemeinschaft (DFG, German 
Research Foundation) through the Collaborative Research Center CRC-TR 211 ``Strong-interaction matter
under extreme conditions'' -- project number 315477589 - TRR 211. E.S.\ acknowledges support by BMBF
``Verbundprojekt: 05P2015 - Alice at High Rate", and 
BMBF ``Forschungsprojekt: 05P2018 - Ausbau von ALICE am LHC (05P18RFCA1)".



\appendix

\section{Lorentz transformation and Jacobian determinant}
\label{app:lorentz}

In this Appendix, we derive Eqs. \eqref{momrel} and \eqref{jacob} in Sec. \ref{sec:momentum}. 
Let $p^\mu=(\varepsilon, {\bf p})$ and $p_*^\mu=(\varepsilon_* , {\bf p}_*)$ be the four-momenta of the 
$\Lambda$ in the laboratory frame and Mother rest frame, respectively, and $P^\mu=(\varepsilon_M, {\bf P})$ 
the four-momentum of the Mother in the laboratory  frame. The pure Lorentz boost transforming the 
momentum of the $\Lambda$ from laboratory to Mother rest frame reads:
\begin{align}\label{lorr1}
\varepsilon_*={}&\gamma (\varepsilon - {\bf V}\cdot {\bf p}), \\
 {\bf p}_* = {}& {\bf p} + \left[\frac{\gamma -1}{\|{\bf V}\|^2}{\bf V} \cdot {\bf p} - 
 \gamma~\varepsilon \right]{\bf V} ,
\label{lorr2}
\end{align}
where ${\bf V}={\bf P}/\varepsilon_M$ is the velocity of the Mother and $\gamma=\varepsilon_M/m_M$ 
the corresponding Lorentz factor. Hence, \eqref{lorr1} and \eqref{lorr2} can also 
be written as:
\begin{align}\label{lorr11}
 \varepsilon_*={}&{1\over m_M} (\varepsilon_M\varepsilon - {\bf P}\cdot {\bf p}) \\
 {\bf p}_* = {}& {\bf p} + \left[ \frac{{\bf P}\cdot{\bf p}}{m_M (\varepsilon_M + m_M)} -\frac{\varepsilon}{m_M} \right]{\bf P} ,
\label{lorr22}
\end{align}
By solving the equation \eqref{lorr11} to get the ${\bf P} \cdot {\bf p}$, we can rewrite 
\eqref{lorr22} as:
\be\label{ppp}
 {\bf p}_* = {\bf p} + \left[\frac{\varepsilon_M\varepsilon - m_M \varepsilon_*}{m_M(\varepsilon_M+m_M)} -  \frac{\varepsilon}{m_M} \right] {\bf P} 
 = {\bf p} - \frac{\varepsilon_* + \varepsilon}{\varepsilon_M+m_M} {\bf P},
\ee
We now move ${\bf p}$ to the left-hand side and take the square 
\begin{align*}
\|{\bf p}_* - {\bf p} \|^2={}& \frac{(\varepsilon_* + \varepsilon)^2}{(\varepsilon_M+m_M)^2} \| {\bf P} \|^2 
\\
={}& \frac{\varepsilon_M-m_M}{\varepsilon_M+m_M}(\varepsilon_* + \varepsilon)^2, 
\end{align*}
where the relativistic dispersion relation has been used in the latter equality. Therefore, the energy 
of the Mother can be obtained by solving the above equation:
\begin{equation}
\label{EEE}
\varepsilon_M=m_M \frac{(\varepsilon_* + \varepsilon)^2 + \|{\bf p}_* - {\bf p} \|^2}{(\varepsilon_* + \varepsilon)^2 
- \|{\bf p}_* - {\bf p} \|^2},
\end{equation}
which, when substituted into \eqref{ppp} yields the momentum of the Mother as a function of the momenta
in both QGP and Mother frames:
$$
{\bf P}= 2m_M \frac{(\varepsilon_* + \varepsilon)({\bf p}-{\bf p}_* )}{(\varepsilon_* + \varepsilon)^2-\|{\bf p}-{\bf p}_* \|^2}.
$$

The above equation is the expression needed to change the integration variable from ${\bf P}$ to 
${\bf p}_*$ by keeping ${\bf p}$ fixed. The Jacobian of the transformation reads 
$$
\frac{\partial {\rm P}_i}{\partial \p_{*j}}=\frac{2m_M}{D^2} \left\{\left[ (\p_{i}-\p_{*i})\frac{\p_{*j}}{\varepsilon_*} 
- (\varepsilon_* + \varepsilon)\delta_{ij} \right]D - 2N_i \left[(\varepsilon_* + \varepsilon)\frac{\p_{*j}}{\varepsilon_*} 
+ (\p_j - \p_{*j}) \right]\right\}
$$
where:
\begin{align*}
 D={}&(\varepsilon_* + \varepsilon)^2-\|{\bf p}_* - {\bf p}\|^2 , \\
 N_i ={}& (\varepsilon_* + \varepsilon)(\p_i-\p_{*i} ).
\end{align*}
After some algebraic manipulations, the determinant of the Jacobian can be computed and 
turns out to be:
$$
\left| \frac{\partial {\bf P}}{\partial{\bf p}} \right|= \frac{m_M^3(\varepsilon_* + \varepsilon)^2 \left[ (\varepsilon_* + \varepsilon)^2 - (\varepsilon \varepsilon_* + {\bf p}\cdot {\bf p}_*+m_\Lambda^2) \right]}{\varepsilon_*(\varepsilon \varepsilon_* + {\bf p}\cdot {\bf p}_*+m_\Lambda^2)^3}
$$
with $m_\Lambda$ the mass of the Daughter, that is Eq.~\eqref{jacob}.

\section{Transferred polarization along angular momentum direction}\label{App:Tpolarization}

Herein, we work out the component of the spin vector of the secondary $\Lambda$ along the angular 
momentum, conventionally opposite to the $y$ axis direction, from \eqref{decayspin2}:
$$
  \langle {\bf S}_{0\Lambda}({\bf p}) \rangle \cdot \hat{\bf j} =
  \frac{\int \di \Omega_* \; n({\bf P})  \left| \frac{\partial {\bf P}}{\partial{\bf p}} \right| \, 
  \left[ S_{*M y} \left( A + B \sin^2 \varphi_* \sin^2 \theta_* \right) + \frac{B}{2} \left( S_{*M z} 
  \sin \varphi_* \sin 2 \theta_* + S_{*M x} \sin 2 \varphi_* \sin^2 \theta_* \right) \right]}
  {\int \di \Omega_* n({\bf P})  \left| \frac{\partial {\bf P}}{\partial{\bf p}} \right| \,},
$$
where we have replaced the expressions of the components of the Mother's spin vector according to 
Eqs. \eqref{harmonics}. As $h_1({\rm P}_T,Y)$ and $g_1({\rm P}_T,Y)$ are odd functions of $Y$,
hence changing sign under the transformation $\theta_*\rightarrow\pi-\theta_*$,  
Eqs. \eqref{harmonics}, \eqref{phimother} and \eqref{coeff} imply that the terms proportional 
to $h_1({\rm P}_T,Y), g_1({\rm P}_T,Y)$ 
vanish after integration in $\theta_*$. Likewise, the term proportional to $f_2({\rm P}_T,Y)$, 
which is an even function under the above transformation, vanishes upon integration over $\theta_*$ 
because it is multiplied by a factor $\sin 2 \theta_*$ which is odd. Therefore, we are left with:
$$
  \langle {\bf S}_{0\Lambda}({\bf p}) \rangle \cdot \hat{\bf j} ={2j(j+1)\over3}
  \frac{\int \di \Omega_* \; n({\bf P})  \left| \frac{\partial {\bf P}}{\partial{\bf p}} \right| \, 
  I(\theta_*,\varphi_*)}
  {\int \di \Omega_* n({\bf P})  \left| \frac{\partial {\bf P}}{\partial{\bf p}} \right| \,},
$$
where the integrand in the numerator is
\begin{align*}
I (\theta_*,\varphi_*) = (g_0 + g_2 \cos 2 \varphi_M) \left( A + \frac{B}{2}(1-\cos 2\varphi_*) 
\sin^2 \theta_* \right) + \frac{B}{2} h_2 \sin 2 \varphi_M\sin^2 \theta_* \sin 2\varphi_*. 
\end{align*}
The term
$$
 g_0\left( A + \frac{B}{2} \sin^2 \theta_* \right)
$$
contributes to the global polarization and is not of special interest here. The remaining part 
of the integrand function $I$, after replacing $\varphi_* = \varphi + \psi$ reads:
\begin{align}\label{trapol2}
 &g_2  \left( A + \frac{B}{2}(1-\cos 2 \varphi \cos 2 \psi + \sin 2 \varphi \sin 2 \psi)  
 \sin^2 \theta_* \right)\cos 2 \varphi_M-\frac{B}{2} g_0 \sin^2 \theta_* \cos 2 \varphi \cos 2 \psi  \nonumber \\   
 & +\frac{B}{2} g_0 \sin^2 \theta_* \sin 2 \varphi \sin 2 \psi +
 \frac{B}{2} h_2 \sin^2 \theta_*  (\cos 2 \varphi \sin 2 \psi + \sin 2 \varphi \cos 2 \psi)\sin 2 \varphi_M . 
\end{align}
The term 
$$
 g_0 \frac{B}{2} \sin^2 {\theta_*} \sin 2 \varphi \sin 2 \psi 
$$
does not contribute after integration as it gives rise to an odd function in $\psi$. 

We can now plug in the \eqref{phimother} to rewrite  \eqref{trapol2} as:
\begin{align*}
 & g_2 \left( A + \frac{B}{2}(1-\cos 2 \varphi \cos 2 \psi + \sin 2 \varphi \sin 2 \psi) \sin^2 \theta_* \right) ({\cal A} \cos 2 \varphi - {\cal B} \sin 2 \varphi) -
 \frac{B}{2} g_0 \sin^2 \theta_* \cos 2 \psi \cos 2 \varphi \nonumber \\   
 & + \frac{B}{2} h_2 \sin^2 \theta_*  (\cos 2 \varphi \sin 2 \psi + \sin 2 \varphi \cos 2 \psi) 
  ({\cal A} \sin 2 \varphi + {\cal B} \cos 2 \varphi). 
\end{align*}
Taking into account that  Eq.~\eqref{psiparity} and that the odd combinations in $\psi$ vanish after 
integration, the effective integrand function is reduced to:
\begin{align*}
 & g_2 \left[\left( A + \frac{B}{2}(1-\cos 2 \varphi \cos 2 \psi) \sin^2 \theta_* \right) {\cal A} \cos 2 \varphi-\frac{B}{2}{\cal B}\sin 2 \psi \sin^2 \theta_* \sin^2 2 \varphi\right] -
 \frac{B}{2} g_0 \sin^2 \theta_* \cos 2 \psi \cos 2 \varphi  \nonumber \\   
 & +\frac{B}{2} h_2 \sin^2 \theta_*  ({\cal A} \cos 2 \psi \sin^2 2 \varphi+ {\cal B}\sin 2 \psi \cos^2 2 \varphi). 
\end{align*}
Finally, by rearranging all the terms and utilizing double-angle trigonometric relationships, we can put the expression in terms of different simple trigonometric functions of $\varphi$:
\begin{align*}
 & \left[g_2 \left( A + \frac{B}{2}\sin^2 \theta_* \right) {\cal A}-
 \frac{B}{2} g_0 \sin^2 \theta_* \cos 2 \psi \right]\cos 2 \varphi + \frac{B}{2} \sin^2 \theta_* {\cal A} \cos 2 \psi (h_2\sin^2 2 \varphi-g_2\cos^2 2 \varphi) \nonumber \\   
 & + \frac{B}{2} \sin^2 \theta_*{\cal B}\sin 2  \psi (h_2\cos^2 2 \varphi-g_2\sin^22 \varphi)\nonumber\\
 =& \left[\frac{B}{4} (h_2-g_2)  \sin^2 \theta_* ({\cal A} \cos 2 \psi+{\cal B}\sin 2  \psi)\right]+\left[g_2 \left( A + \frac{B}{2}\sin^2 \theta_* \right) {\cal A}-
 \frac{B}{2} g_0 \sin^2 \theta_* \cos 2 \psi \right]\cos 2 \varphi  \nonumber \\   
 & - \left[\frac{B}{4} (h_2+g_2) \sin^2 \theta_*({\cal A} \cos 2 \psi-{\cal B}\sin 2  \psi)\right]\cos 4 \varphi.
\end{align*}
The first term contributes to the global polarization, whereas the second and the third are the
expected leading azimuthal modulations of this component.

\end{document}